\documentclass[conference]{IEEEtran}
\IEEEoverridecommandlockouts
\usepackage{cite}
\usepackage{amsmath,amssymb,amsfonts}
\usepackage{algorithmic}
\usepackage{graphicx}
\usepackage{textcomp}
\usepackage{tabularx,booktabs}
\usepackage{xcolor}
\def\BibTeX{{\rm B\kern-.05em{\sc i\kern-.025em b}\kern-.08em
    T\kern-.1667em\lower.7ex\hbox{E}\kern-.125emX}}
\usepackage{lipsum}
 
\begin{document}
\title{Detecting Synthetic Speech Manipulation in Real Audio Recordings\\}

\author{\IEEEauthorblockN{Md Hafizur Rahman}
\IEEEauthorblockA{STAR Lab \\SRI International \\ \texttt{mdhafizur.rahman@sri.com}}
\and
\IEEEauthorblockN{Martin Graciarena}
% our CI requires the university name to be written in full length...
\IEEEauthorblockA{STAR Lab \\SRI International \\ \texttt{martin.graciarena@sri.com}}
\and
\IEEEauthorblockN{Diego Castan}
\IEEEauthorblockA{STAR Lab \\SRI International \\ \texttt{diego.castan@sri.com}}
\and
\IEEEauthorblockN{Chris Cobo-Kroenke}
\IEEEauthorblockA{STAR Lab \\SRI International \\ \texttt{chris.cobo-kroenke@sri.com}}
\and
\IEEEauthorblockN{Mitchell McLaren}
\IEEEauthorblockA{STAR Lab \\SRI International \\ \texttt{mitchell.mclaren@sri.com}}
\and
\IEEEauthorblockN{Aaron Lawson}
\IEEEauthorblockA{STAR Lab \\SRI International \\ \texttt{aaron.lawson@sri.com}}
}

\maketitle

\begin{abstract}
Recent advances in artificial speech and audio technologies have improved the abilities of deep-fake operators to falsify media and spread malicious misinformation. Anyone with limited coding skills can use freely available speech synthesis tools to create convincing simulations of influential speakers' voices with the malicious intent to distort the original message. With the latest technology, malicious operators do not have to generate an entire audio clip; instead, they can insert a partial manipulation or a segment of synthetic speech into a genuine audio recording to change the entire context and meaning of the original message. Detecting these insertions is especially challenging because partially manipulated audio can more easily avoid synthetic speech detectors than entirely fake messages can. This paper describes a potential partial synthetic speech detection system based on the x‑ResNet architecture with a probabilistic linear discriminant analysis (PLDA) backend and interleaved aware score processing. Experimental results suggest that the PLDA backend results in a 25\% average error reduction among partially synthesized datasets over a non-PLDA baseline.

\end{abstract}

\begin{IEEEkeywords}
partial synthetic speech, synthetic speech detection, anti-spoof, x-ResNet, PLDA
\end{IEEEkeywords}

\section{Introduction}
In the last decade, the rise of artificial intelligence (AI)-driven deep-fake technologies has become a major concern in society. Some recent reports suggest that widespread misinformation and fraud are already taking place primarily in the financial and political sectors \cite{ice2019defamatory}. %Anyone who can access open-source or cheaply available commercial audio deep-fake tools can imitate the voice of any target speaker to produce malicious content to defame or breach security. 
Audio deep-fake technologies can be broadly categorized into two types: text-to-speech (TTS) and voice-cloning (VC). While TTS systems generate synthetic speech from a text input that sounds like the target speaker, VC systems maintain the source speaker's linguistic content. Although these technologies are not new to the community, audio deep-fake entered a new era after DeepMind released Wavenet \cite{oord2016wavenet} to synthesize almost real human voices. Some recent TTS and VC conversion tools, including DeepVoice \cite{arik2017deep}, Fast Speech 1 \& 2 \cite{ren2019fastspeech,ren2020fastspeech}, Flowtron \cite{valle2020flowtron}, Tacotron 2 \cite{shen2018natural}, FastPitch \cite{lancucki2021fastpitch}, and RAD-TTS \cite{shih2021rad}, can produce high quality synthetic speech that is almost impossible to distinguish by human ear and increases the need for robust fake or synthetic speech detectors.

Synthetic audio detection is a very challenging problem. To address spoof attacks on the automatic speaker verification (ASV), the research community introduced the ASV spoof challenges \cite{wu2015asvspoof,kinnunen2017asvspoof,todisco2019asvspoof}. 
The spoof challenges generated studies of two major ASV spoof attacks: (1) logical access (LA) or synthetic speech attack (VC and TTS systems) and (2) physical access (PA) or replay attack. Although the ASVspoof challenge paved the way for synthetic audio detection research, these datasets include only fully synthetic audio. However, partial manipulation of speech audio can change the entire meaning of the media content and, currently, is being used to spoof security systems or defame important public figures.   
In spite of the risks created by these attacks, very few studies are being conducted in this area. We found only one English partially synthetic speech dataset called PartialSpoof \cite{zhang2021initial} based on the ASV spoof 2019 dataset, and one Mandarin partially synthetic speech dataset called Half-Truth audio dataset (HAD) \cite{yi2021half}. 

The research community explored strategies to address audio spoofing, primarily with different front-end spectro-temporal features and statistical and deep learning modeling approaches. Pristine and generated speech-discriminant information residing in the spectro-temporal domain can be beneficial in modeling robust synthetic audio detectors. Popular features like Linear Frequency Cepstral Coefficients (LFCC) , constant-Q cepstral coefficients (CQCC) [15], modified group delay function (MGDF) \cite{wu2012study,wu2012detecting}, and Cosine Phase function (cosphase) \cite{wu2012detecting} have shown promise by achieving low errors in ASVspoof challenges. However, the cross-dataset evaluation shows that those hand-crafted features cannot capture sufficient variability between datasets to train back-end classifiers \cite{das2020assessing}. More recent studies incorporated Deep Neural Network (DNN) based deep feature extractor strategy for better generalization \cite{wang2021investigating,teng2021complementing}. For synthetic speech detection, a statistical model like Gaussian Mixture Models (GMM) is regarded as the baseline system [21], while deep learning frameworks including LCNN \cite{lavrentyeva2019stc}, SENet \cite{wu2020defense}, VGG \cite{wu2020defense}, ResNet \cite{chen2017resnet}, Res2Net \cite{li2021replay}, and more recently, end-to-end systems like RawNet2 \cite{tak2021end} and RW-ResNet \cite{ma2021rw} showed good progress in this challenge. A few studies focused on deep learning training loss functions, including one-class (OC) loss and \cite{zhang2021one}
large margin cosine loss (LMCL) \cite{chen2020generalization}, which are particularly successful in improving DNN generalization. However, most individual systems fail to produce good performance across all the tasks or datasets. Ensemble or fused systems showed good overall performance, but these systems are unsuitable for real-world application due to training and run-time computational complexities.

In this paper, we present a synthetic speech detection system based on x-ResNet architecture with a PLDA back-end classifier. The x-ResNet is a variation of ResNet architecture with some modifications in the down-sampling blocks to utilize more discriminant information from the input features \cite{he2019bag}. In the back-end, we leveraged the PLDA \cite{prince2007probabilistic} classifier, which typically provides better generalization across data conditions and is widely used by the speaker and language recognition research community. We also introduce several very challenging partially synthetic audio deep-fake datasets for this task. Detecting synthetic audio manipulation or audio insertion in a pristine audio is a very challenging task. For partially synthetic audio detection we introduce interleaved aware score post-processing based on window-score smoothing.

The remainder of this paper is structured as follows: Section 2 details the synthetic speech detection system used in this paper. Section 3 outlines the experimental setup, datasets, and results. Finally, Section 4 presents conclusions.

\begin{figure*}[!t]
  \includegraphics[width=\textwidth,height=3.95cm]{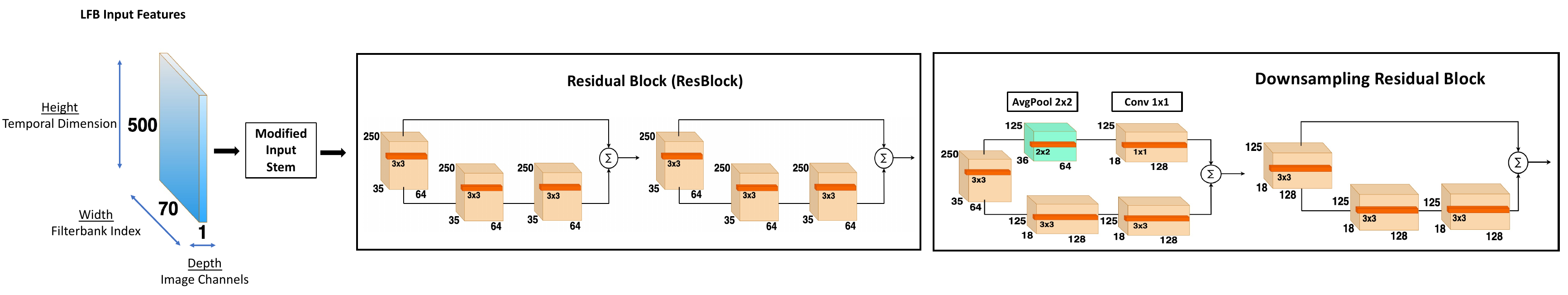}  
  \caption{The x-ResNet-18 network with input stem and example Residual blocks. The x-ResNet has a modified input stem compared to its original version. The modified input stem consists of three 3x3 convolutional layers instead of the original 7x7 convolutional layer. The first Residual stage always passes through the full-sized input. After the first stage, every other stage down samples the input and doubles the number of filters to keep the computation constant.}
  \vspace{-3mm}
\end{figure*}

\section{System Description}
The overall system is composed of three main parts: the audio front-end for feature extraction, deep learning model architecture, and the PLDA back-end for scoring.

\subsection{Audio Front-end}
%The front-end part of the system is composed of the acoustic features, and the speech activity detector (SAD).

\subsubsection{Acoustic Features}
Linear Filter Banks (LFB) is an acoustic feature extraction technique that uses a series of filter banks on a linear frequency scale (uniform frequency separation between filters), which is computationally less expensive than constant-Q and other cepstral features. LFB obtains a higher spectral resolution at high frequencies because the separation between filters does not increase at higher frequencies like Mel Filter Bank (MFB) energies. The motivation behind adding full frequency coverage is to expose the system to artifacts in synthetic speech left behind by speech generators that may be in frequency regions outside the fixed spectral range of human speech. We used 70 triangular linearly spaced filters to extract LFB features from a 25 ms window with a 10 ms frameshift.

\subsubsection{Speech Activity Detection (SAD)}
One recent study \cite{muller2021speech} found that silent segments in the training data have a significant impact on the ASV spoof countermeasure performance. Although silences seem beneficial to model training, but then DNNs fail to produce good generalization performance across datasets when trained with pauses and silences. One potential reason is that critical information for generalization is overwhelmed (or suppressed) by the presence of perceivably easy silence information. To avoid this scenario, we applied Speech Activity Detector (SAD) in the study based on a feed-forward DNN architecture. Our DNN model is trained with 20-dimensional Mel Frequency Cepstral Coefficients (MFCC) features with a 0.31 second long input temporal window and 0.5 second long mean and variance normalization window. The DNN model topology is two-layers with 500 and 100 neurons in each layer, respectively. The output scores are smoothed with a 0.5 second long window. Finally, the detected speech segments are padded by a third of a second at both edges.

\subsection{Deep-learning Model Architecture}
\subsubsection{X-ResNet}
The x-ResNet architecture is based on regular ResNet with minor modifications in the down sampling block. The ResNet architecture was first proposed to address the vanishing gradient \cite{he2016deep} problem of the deeper networks by adding identity skip-connections between layers. This network consists of an input stem followed by four Residual stages and an output layer. Each stage of the ResNet consists of several Residual blocks depending on how deep the model is. From stage 2, a down sampling block is added at the beginning of each stage. The x-ResNet variation studied in this work first appeared in \cite{he2019bag} and has been successfully applied on image classification tasks. The main idea behind this modification is to attain more usable information in the down sampling blocks discarded in the regular down sampling block.

Figure 1 shows the parts of the x-ResNet network in detail with example Residual blocks. This x-ResNet network has a modified input stem compared to its original version. The original 7x7 convolutional layer is replaced with three 3x3 convolutional layers with stride 2 in the first layer for down sampling, with 32 filters in the first two layers, and 64 filters in the last layer. This is computationally equivalent to the original, with more expressive power from deeper layers. 
After the input stem, a typical ResNet consists of four Residual stages. Each stage has a certain number of Residual blocks. The number of blocks per stage determines the final depth of the network. The first Residual stage always passes through the full-sized input. After the first, every other stage down samples the input and doubles the number of filters to keep the computation constant. The down sampling is performed with 2x2 average pooling for its anti-aliasing benefits. The 1x1 convolution afterward increases the number of feature maps, matching the residual path's output. The final output of the Residual stages is then fed to the statistical pooling and embedding layers for further processing. Embeddings are extracted from the last layer of the x-ResNet before the final soft-max layer.
  \vspace{-1mm}
%To extract embeddings from the xResNet, we compute the mean of the last layer of the xResNet before the final output.
\subsubsection{SE Block}
Squeeze-and-Excitation (SE) \cite{hu2018squeeze} blocks are very efficient in adaptively re-calibrating convolution channel inter-dependencies into a global feature so that the dominant channels can achieve higher weights, which typically leads to a performance gain. The squeeze operation uses global average pooling followed by generating each independent channel's weights during the excitation operation. The excitation operation is performed with a bottleneck architecture, where the first layer reduces dimension with a Rectified Linear Unit (ReLu) activation followed by extending the back dimension with sigmoid activations in the next layer. The output of the sigmoid activations is used as channel weights with a higher emphasis on the most significant channels.

\subsubsection{OC-Softmax}
Zhang et al. \cite{zhang2021one} introduced a one-class feature learning approach to train a compact embeddings space by introducing margins to consolidate target real speech and isolate synthetic speech data to prevent ResNet models from over-fitting to any seen speech generators. In our investigation, we applied the OC-Softmax function for model training to improve DNN generalization across unknown conditions using following equation,

\begin{equation}
L_{OC} = \frac{1}{N}\sum_{i=1}^{N}{\log(1+e^{{\alpha}(m_{y_{i}}-\mathbf{\hat{w}_{0}}\mathbf{\hat{x}_{i}})(-1)^{y_{i}}}})
\vspace{-1mm}
\end{equation}
where $\mathbf{\hat{x}_{i}} \in \mathbb{R}^{D}$ and  $\mathbf{\hat{w}_{0}} \in \mathbb{R}^{D}$ represents the normalized target-class embeddings and weight-vector, respectively. $y_{i} \in {0, 1}$ denotes sample labels, and $m_{0}$, $m_{1} \in[-1, 1]$, where $m_{0} > m_{1}$ are the angular margins between classes. 

\subsection{Data Augmentation}
%We don't want to train our model to latch too much into the silences of the audio.
One of the key reasons for poor deep-learning model performance is model over fitting. There are very few publicly available datasets for deep-fake model training. Most of the available datasets also come with a limited number of speech generators. To minimize the chances of model over fitting and to improve generalization, we applied different data augmentation strategies. We applied four different types of audio degradation into our training set: (1) reverb, (2) compression, (3) instrumental music, and (4) noise, which includes babble; restaurant; different indoor and outdoor sounds; traffic; mechanical; and other natural noises. We then augmented the training set with these four types of degradation at a 5-dB signal-to-noise (SNR) ratio, a random selection of noises, a random selection of 30 different bit-rate compressions, and a random selection of reverberated signals with low reverberation. More details about the data augmentation strategy can be found in \cite{mclaren2018train}.
We also applied the frequency masking \cite{chen2020generalization} technique to randomly dropout frequency bands during training ranging from $f_{0}$ to $f_{0}+f$, where $f$ is chosen from a uniform distribution from $0$ to a maximum number of masked channels, $F$. Different frequency masks with randomly chosen sizes between one and seven pixels were applied to each sample in a batch. The masked spectral regions were filled in with the average value of the entire feature. 
\subsection{Back-end}

\subsubsection{Probabilistic Linear Discriminant Analysis (PLDA)}
In this task, we leveraged a PLDA-backend scorer for x-ResNet embeddings systems to compute scores between enrol and test embeddings. Embeddings are extracted after removing the output layer from the network. 
The PLDA backend starts by reducing the dimension with LDA transformation and gaussianization of the input embeddings $\mathbf{w_{i}}$, which can be modelled as,
\begin{equation}
\mathbf{w_{i}}={\mathbf{\mu} + U_{1}\cdot x_{1} + \mathbf{\epsilon_{i}}}
\end{equation} 
where $\mathbf{\mu}$ is the mean vector, $U_{1}$ is the eigen matrix, $x_{1}$ is the hidden factor, and $\mathbf{\epsilon_{i}}$ models the residual variability.

%Given a pre-prosessed enrol $w_{e}$ and test embeddings $w_{t}$ the PLDA scores are computed with following log-likelihood-ratio (LLR),
%\begin{equation}
%s = log\frac{P(w_{e},w_{t}|H_{s})}{P(w_{e},w_{t}|H_{d})}
%\end{equation}
%
%where, $H_{s}$ and $H_{d}$ are the hypotheses that the enrollment and test samples from same or different models, respectively.

\subsubsection{Interleaved Aware Score Post-processing}
%Most of the spoof countermeasure systems in the literature either apply a frame/chunk score averaging approach or produce one score per utterance by randomly selecting speech chucks from the audio, considering the audio is either entirely pristine or entirely generated. These approaches are not very reliable for detecting minor inconsistencies in the audio. Instead, we applied simple score post-processing based on window-score smoothing. The output segmental scores from the PLDA backend are smoothed with a multiple windows mean filter of length ten frames. Finally, the average score of the top 5\% smoothed window scores.

One of the ways to estimate an utterance level score from the segment scores is to average all output segment scores to get a single score to represent the whole waveform. However, this approach is not very reliable for detecting small inconsistencies in the audio. Instead, we applied simple interleaved aware score post-processing based on window-score smoothing. The output segmental scores from the PLDA backend are smoothed with a multiple windows mean filter of ten frames. Finally, the average score of the top 5\% smoothed window scores is estimated as the final score for the whole waveform.
%\newpage

\section{Experiments}
\subsection{Setup}
We extracted 70 dimensional LFB features from the raw audio and applied a SAD mask on the speech features to remove silent portions from the audio. x-ResNet models are trained with frequency masked, and four copies of noise-augmented data as mentioned in Section 2.3. The x-ResNet was trained with a batch size of 64 samples over 20 epochs using the OCSoftmax loss function. The Adam optimizer was used with a learning rate of 2e-3, beta terms of (0.9, 0.99), and an epsilon term of 1e-5. Weight decay was also applied with a scaling factor of 0.01. The training was stopped after 12 epochs without reductions in the equal error rate (EER) on the validation set.
We extracted 64-dimensional embeddings from a 500 frames sliding window with 10 frames shift and from the last layer of the network after stripping the output layer. The PLDA subspace was trained with embeddings from 10 pristine and 10 balanced speech generator classes. 
%Frequency masking was applied to the input feature in the form of SpecAugment. Different frequency masks with randomly chosen sizes between one and seven pixels were applied to each sample in a batch. The masked spectral regions were filled in with the average value of the entire feature.
\subsection{Datasets}
\subsubsection{Training}
Our training pool includes genuine and synthetic speech from a diverse collection of state-of-the-art TTS and VC algorithms for better DNN generalization and prevents the model from over fitting to any known speech generators or datasets. We used 17 speech generators' samples from the training subset of the ASVspoof 2019 \cite{todisco2019asvspoof} logical attack (LA) task, which includes 2,580 real and 22,800 generated speech utterances. Since ASVspoof samples are relatively smaller in duration, we used half of the generated data from each speech generator to avoid duration mismatch during the model training. Another publicly available synthetic speech data collection called Fake-or-Real (FoR) \cite{reimao2019dataset} contains more than 111,000 bonafide and 87,000 synthetic speech samples from 33 different speech generators. We used 53,862 pristine and 26,924 synthetic speech samples from FoR training set.
In addition, we generated 30,000 TTS samples from RTVC \cite{jemine2019master} and Tacotron2 \cite{shen2018natural} generators trained with 80 LibriTTS speakers. From pristine audio datasets like LJSpeech and  LibriTTS, we used 9,988 and 21,317 pristine samples, respectively.

\subsubsection{Evaluation}
We used two groups of datasets for synthetic speech detection evaluation. The first group includes the ASVspoof2019 LA dataset and the FoR dataset, which includes only fully pristine and fully generated audio samples. The ASVspoof2019 LA evaluation dataset contains 7,355 real samples from 67 speakers and 63,882 generated samples from 48 speakers. %Generated samples are produced using various state-of-the-art TTS syntheses and VC algorithms.
The FoR evaluation set comes with gender and speaker balanced 2,264 pristine and 2,370 generated samples from Google TTS Wavenet \cite{oord2016wavenet}.

The second group contains partially synthetic speech samples from the PartialSpoof database, Semantic Forensics (SemaFor) internal program evaluation 1, SemaFor hackathon2 data, and in-house generated partially synthetic datasets. The PartialSpoof \cite{zhang2021initial} (PS) database has been derived from the ASVspoof 2019 LA challenge by randomly inserting spoof audio segments in pristine audio waveforms. This database consists of 22,296  samples in the development set and 63,882 samples in the evaluation set. The evaluation set includes 0.5–2 seconds of inserted synthetic audio out of 3-5 seconds of audio clips. SemaFor is a DARPA-funded research program that aims at developing semantic technologies for analyzing multimedia to counter deep-fake technologies. SemaFor eval1 (SF-EV1) is a small, partially synthetic test set produced for this program to evaluate generated audio detection systems. This test set consists of 172 pristine and 173 partially synthetic samples produced from a single speaker with RTVC and Tacotron2 generators. Test samples consist of an average of 1 minute of synthetic manipulation out of 3-4 minutes of full-length audio. SemaFor hackathon2 (SF-HK2) is a more challenging dataset produced with the latest speech generators like RAD-TTS and Tacotron2 (Jarvis) by Nvidia. Audio samples are heavily degraded with white-noise, factory, music, and riot noises. This dataset consists of 384 pristine and 1024 partially synthetic samples from 4 speakers. This dataset consists of 4-5 seconds of synthetic manipulation out of 10-15 seconds of full audio. Our in-house generated (IG) partially synthetic test set is prepared by randomly inserting ~1-3 seconds long generated samples into the pristine samples. This test set consists of 45,566 pristine and 33,275  partially synthetic samples from 250 speakers. Generated samples are produced with RTVC \cite{jemine2019master} and Tacotron2 \cite{shen2018natural} TTS. This partially synthetic test set consists of 2-3 seconds of audio manipulation out of 20-30 seconds of full audio.

\subsection{Results}

\begin{table}[t]
\caption{EER (\%) performance comparison of different systems on fully generated (LA, FoR) and partially synthetic (PS, SF-EV1, SF-HK2, and IG) datasets .}
	
\begin{tabular}{c|c|c|c|c|c|c}

		\toprule
		{System}	& {LA}		&	{FoR}	& {PS}	&	{SF-EV1} 	& {SF-HK2} & {IG} \\
		\midrule
		\multicolumn{7}{c}{Score Averaging} \\
		GMM			&	22.65	&	18.14	&	25.61	&	8.45		&	40.56		&	36.32\\
		ResNet		&	10.42	&	4.58	&	15.45	&	6.81		&	23.18		&	22.03\\
		x-ResNet		&	7.63	&\bf 1.89	&	12.40	&	5.86		&	21.38		&	21.71\\
		SE-xResNet 	&\bf 6.28	&	9.56	&	14.18	&	3.69		&	35.63		&	26.03\\
		\midrule	
		
		\multicolumn{7}{c}{Interleaved aware processing} \\
		GMM			&	26.32	&	31.63	&	26.32	&	1.53		&	38.92		&	29.69	\\
		ResNet 		&	12.58	&	8.56	&	15.10	&	1.28		&	18.2  		&	15.31	\\
		x-ResNet		&	9.51	&	3.8		&	14.23	&\bf 0.58		&\bf 14.06	 	&	16.40	\\
		SE-xResNet	&	8.22	&	18.44	&{\bf 11.98}	&\bf 0.58		&	27.34		&\bf 14.39	\\
		\bottomrule
	\end{tabular}

	\label{tab:tabdnn}

\end{table}

\begin{table}[t!]
	\centering

	\caption{EER (\%) performance comparison of different systems on fully generated (LA, and FoR) and partially synthetic (PS, SF-EV1, SF-HK2, and IG) datasets with embeddings PLDA backend.}
	\begin{tabular}{c|c|c|c|c|c|c}
		\toprule
		{\bf System}& {LA}		& {FoR}		& {PS}	& {SF-EV1} & {SF-HK2} 	& {IG}\\
		\midrule
		\multicolumn{7}{c}{Score Averaging} \\
		ResNet 		&	9.92	&	5.1		&	16.82	&	4.1			&	28.14  		&	18.63	\\
		x-ResNet		&\bf 6.88	&	3.56	&	13.78	&	3.12		&	18.61		&	16.67\\
		SE-xResNet	&	7.51	&	2.10	&	12.45	&	4.10		&	31.87		&	14.74\\
		\midrule
		
		\multicolumn{7}{c}{Interleaved aware processing} \\
		ResNet 		&	11.21	&	8.92	&	13.22	&	0.58		&	21.74  		&	14.23	\\
		x-ResNet		&	8.91	&\bf 2.60	&	10.61	&	\bf 0.00	&	\bf 11.52   &	\bf 10.27	\\
		SE-xResNet	&	8.95	&	5.32	&\bf 10.58	&	0.58		&	25.52   	&	12.89	\\
		\bottomrule
	\end{tabular}
	\label{tab:tabplda}

\end{table}

Table \ref{tab:tabdnn} shows the EER performance of the systems evaluated on fully generated and  partially synthetic datasets. This experimental setup used two baseline systems, GMM and ResNet, with OC loss function for performance comparison. Experimental results show that both x-ResNet and SE-xResNet systems performed significantly better than baseline systems. On average, we found at least 54\% and 21\% relative error reduction across the datasets with the x-ResNet compared to the baseline GMM and ResNet systems, respectively. In addition, backend score processing played an important role, especially in partially synthetic audio detection. While interleaved aware score processing showed a considerable increase in error in the fully generated ASVspoof LA and FoR datasets, but produced a 33.5\% error reduction on partially spoofed audio datasets. 
The SF-HK2 dataset was the most difficult among all  partially synthetic datasets since test samples are heavily degraded with external noises, and the x-ResNet system achieved the best performance of 14.06\% EER.
With interleaved aware score processing, the SE-xResNet system performed well on partially spoofed datasets, but on average the x-ResNet system performed relatively well across all datasets with the non-PLDA backend.

The performance of the synthetic audio detection systems with embeddings PLDA backend are presented in Table \ref{tab:tabplda}. 
In most cases, we observed a good error reduction with the PLDA backend compared to the raw DNN scores. Like non-PLDA systems, x-ResNet systems with embeddings PLDA backend showed a 25\% reduction in error with interleaved aware score processing across partially generated datasets. In Table \ref{tab:tabdnn}, the loss in performance due to interleaved aware processing on fully generated datasets is quite significant. However, the PLDA backend helped remove that separation. 
The SE-xResNet did not provide any additional gains across the board, as observed in Table \ref{tab:tabdnn}, but it produced the best performance of 10.58\% EER on the PartialSpoof dataset. SF-EV1 dataset is a relatively easier dataset among other partially synthetic datasets since both generators are known to the model, and all samples are generated from a single speaker. Also, the abrupt injection of 1-3 seconds samples into the speech may not recreate a realistic scenario. Both x-ResNet and SE-xResNet systems performed well on this dataset, and the x-ResNet system produced 0\% EER. It is feasible that our approach hinges on these abrupt signals to score well. Perhaps detectors will struggle if the signal injected is made to sound more natural and the length of the inserted signal is extremely short (less than 0.5 seconds).

\section{Conclusions}
In this paper, we explored synthetic audio detection on partially synthetic audio, which is considered a more challenging and realistic scenario than detecting fully synthetic audio. We explored x-ResNet architecture in combination with SE block and embeddings PLDA backend for the detection countermeasure. We tested our systems on fully and partially generated speech datasets. Experimental results showed that interleaved aware score processing helps improve overall performance on the partially generated datasets while losing some performance on fully generated datasets. The x-ResNet with a PLDA backend helps bridge that gap and achieved at least 25\% error reduction across partially generated datasets. In our future work, we will create a more realistic partially generated dataset for the community, in which instead of abruptly injecting generated speech into real audio, we will replace a few words or sentences in the audio with synthetic samples from the same speaker to make them more realistic. The proposed synthetic audio detection approach can determine if audio is manipulated with synthetic insertion but can not determine the location of the synthetic portion of the audio. In future work, we will address synthetic audio localization in real audio recordings.

\section*{Acknowledgement}
Distribution Statement "A" (Approved for Public Release, Distribution Unlimited). This research was developed with funding from the Defense Advanced Research Projects Agency (DARPA). The views, opinions and/or findings expressed are those of the author and should not be interpreted as representing the official views or policies of the Department of Defense or the U.S. Government.

\bibliographystyle{IEEEtran}
\bibliography{mybib}

\end{document}